\documentclass[twocolumn]{aastex63}
\usepackage{amsmath}
\usepackage{amssymb}
\usepackage{graphicx}
\usepackage{color}
\usepackage{natbib}
\def\gtaprx {\lower .1ex\hbox{\rlap{\raise .6ex\hbox{\hskip .3ex
	{\ifmmode{\scriptscriptstyle >}\else
		{$\scriptscriptstyle >$}\fi}}}
	\kern -.4ex{\ifmmode{\scriptscriptstyle \sim}\else
		{$\scriptscriptstyle\sim$}\fi}}}
\def\ltaprx {\lower .1ex\hbox{\rlap{\raise .6ex\hbox{\hskip .3ex
	{\ifmmode{\scriptscriptstyle <}\else
		{$\scriptscriptstyle <$}\fi}}}
	\kern -.4ex{\ifmmode{\scriptscriptstyle \sim}\else
		{$\scriptscriptstyle\sim$}\fi}}}

\newcommand{\cutt}[1]{\textcolor{blue}{}}

\newcommand{\Ms}{{\ensuremath{M_{\odot} }}}

\shorttitle{UHZ1}
\shortauthors{Whalen et al.}

\begin{document}

\title{Radio Emission From a $z =$ 10.1 Black Hole in UHZ1}

\author{Daniel J. Whalen}

\affiliation{Institute of Cosmology and Gravitation, Portsmouth University, Dennis Sciama Building, Portsmouth PO1 3FX}

\author{Muhammad A. Latif}

\affiliation{Physics Department, College of Science, United Arab Emirates University, PO Box 15551, Al-Ain, UAE}

\author{Mar Mezcua}

\affiliation{Institute of Space Sciences (ICE, CSIC), Campus UAB, Carrer de Magrans, 08193 Barcelona, Spain}
\affiliation{Institut d'Estudis Espacials de Catalunya (IEEC), Carrer Gran Capit\`{a}, 08034 Barcelona, Spain}

\begin{abstract}

The recent discovery of a 4 $\times$ 10$^7$ \Ms\ black hole (BH) in UHZ1 at $z =$ 10.3, just 450 Myr after the big bang, suggests that the seeds of the first quasars may have been direct-collapse black holes (DCBHs) from the collapse of supermassive primordial stars at $z \sim$ 20.  This object was identified in {\em James Webb Space Telescope} ({\em JWST}) NIRcam and {\em Chandra} X-ray data, but recent studies suggest that radio emission from such a BH should also be visible to the Square Kilometer Array (SKA) and the next-generation Very Large Array (ngVLA).  Here, we present estimates of radio flux densities for UHZ1 from 0.1 - 10 GHz, and find that SKA and ngVLA could detect it with integration times of 10 - 100 hr and just 1 - 10 hr, respectively.  It may be possible to see this object with VLA now with longer integration times.  The detection of radio emission from UHZ1 would be a first test of exciting new synergies between near infrared (NIR) and radio observatories that could open the era of $z \sim$ 5 - 15 quasar astronomy in the coming decade.

\end{abstract}


%\maketitle

\section{Introduction}

Nearly 300 quasars have now been discovered at $z >$ 6 \citep{fbs23}, including nine at $z > 7$ \citep{mort11,ban18,yang20,wang21}.  Both DCBHs and BHs forming from the collapse of normal Population III (Pop III) stars have been proposed as the seeds of these quasars. Pop III BHs are thought to form at $z \sim$ 25 with masses of a few tens to hundreds of solar masses and must accrete continuously at the Eddington limit or grow by super-Eddington accretion to reach 10$^9$ \Ms\ by $z \sim$ 7 \citep{vsd15,inay16}.  This scenario is problematic because Pop III BHs are born in low densities that prevent rapid initial growth \citep[e.g.,][]{wan04,latif22a} and can be ejected from their host halos (and thus their fuel supply) during collapse \citep{wf12}.  If they begin to accrete later, they again drive gas out of their host halos because the shallow gravitational potential wells of their host halos, so they are restricted to fairly low duty cycles \citep{srd18}.  

In contrast, DCBHs from the collapse of more rare 10$^4$ - 10$^5$ \Ms\ primordial stars \citep{hos13,um16,tyr21a,herr23a} at $z \sim$ 20 can grow at much higher rates at birth because Bondi-Hoyle accretion rates scale as $M_{\mathrm{BH}}^2$. They also form in much higher ambient densities \citep[e.g.,][]{pat23a} in more massive cosmological halos that retain gas even when it is heated by X-rays (\citealt{latif21a} - see \citealt{titans} for reviews).  However, it has been thought until recently that DCBH formation required unusual or exotic environments that had little chance of coinciding with the rare, massive, low-shear halos needed to fuel the growth of the BH to 10$^9$ \Ms\ by $z \sim$ 7 \citep{ten18,lup21,vgf21}.  It is now known that the highly supersonic turbulence in these rare gas reservoirs can create DCBHs without the need for strong UV backgrounds, supersonic baryon streaming motions, or even atomic cooling \citep{latif22b}.  Supermassive stars could be detected in the near infrared (NIR) at $z \sim$ 8 - 14 \citep{sur18a,sur19a,vik22a,nag23} and DCBHs could be detected at $z \gtrsim$ 20 in the NIR \citep{nat17,bar18,wet20b} and at $z \sim$ 8 - 10 in the radio \citep{wet20a,wet21a}.

\begin{figure} 
\plotone{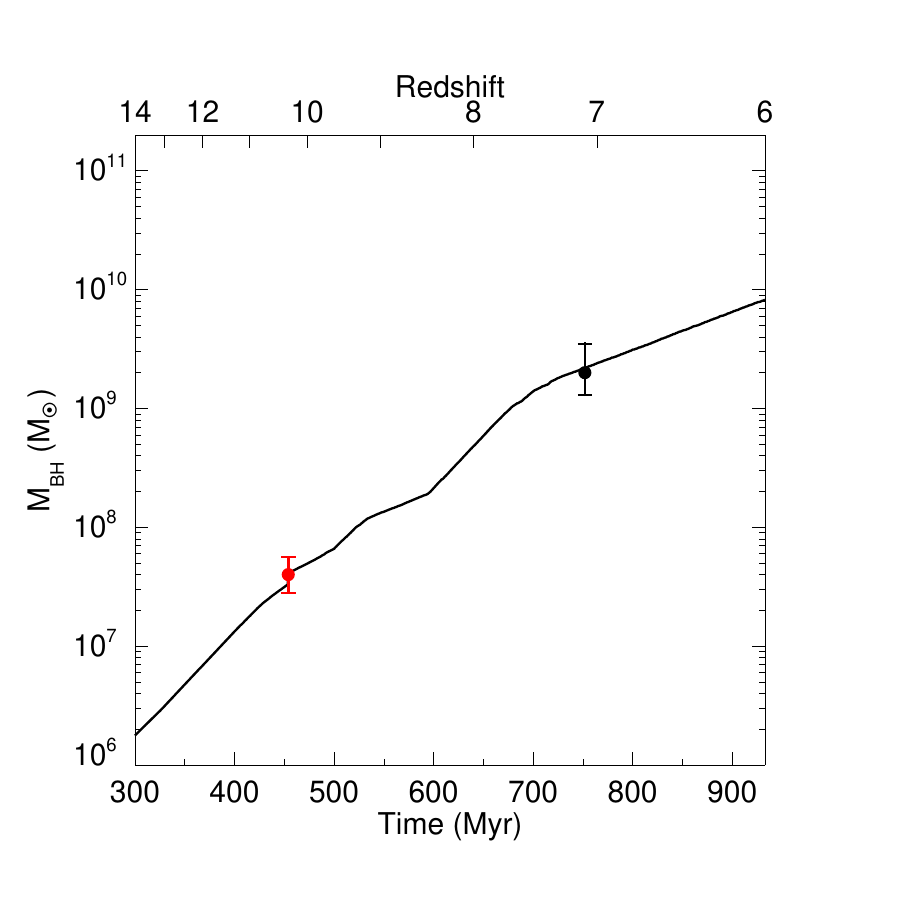}
\caption{BH mass from \citet{smidt18}, who modeled the formation of a quasar that is similar to  J1120 (black), a 1.35 $\times$ 10$^9$ \Ms\ BH at $z =$ 7.1.  UHZ1 (red), with a mass of 4 $\times$ 10$^7$ \Ms, lies directly on this growth curve at $z =$ 10.3 and may be a $z \sim$ 7 quasar at an earlier stage of evolution.}
\vspace{0.1in}
\label{fig:bhmass} 
\end{figure}

Observations of quasars at earlier stages of evolution, above $z \sim$ 7, are clearly needed to distinguish between these formation pathways. One such object may have now been found, a $4 \times 10^7$ \Ms\ BH in UHZ1, which has been identified in {\em JWST} NIRcam and {\em Chandra} X-ray images \citep[RA=0:14:16.096, Dec=-30:22:40.285;][]{cet23,akos23}.  UHZ1 is a small galaxy now spectroscopically confirmed to be at $z =$ 10.1 \citep{gould23} that is gravitationally lensed by the Abel 2744 cluster (magnification $\mu =$ 3.8) and has a star formation rate (SFR) of 4 \Ms\ yr$^{-1}$.  \citet{akos23} suggest that the large mass of the BH at such early times favors a DCBH seed, which has now been corroborated by fits to theoretical spectra for DCBHs residing in overmassive black hole galaxies \citep[OBGs;][]{nat23}.  This result is also consistent with \citet{smidt18}, who followed the growth of a 10$^5$ \Ms\ DCBH at $z \sim$ 19 into a quasar similar to J1120, a $1.35 \times 10^9$ \Ms\ BH at $z = 7.1$ \citep{mort11}, in a cosmological simulation with radiation hydrodynamics.  As shown in Figure~\ref{fig:bhmass}, the BH in UHZ1 falls on the growth curve of that quasar.  Recent calculations indicate that radio emission from this quasar should be visible to SKA and ngVLA at up to $z \sim$ 14 - 16 \citep{latif23a}. Here, we estimate the radio flux density expected from UHZ1 to determine if it could be detected and unambiguously confirm the presence of a BH there.  In Section 2 we lay out our calculation of radio emission from the BH and H II regions and supernovae (SNe) in its host galaxy.  We discuss prospects for its detection in current and future surveys in Section 3 and conclude in Section 4.

\section{Numerical Method}

We calculate radio flux densities for the BH in UHZ1 from 0.1 - 10 GHz with fundamental planes (FPs) of BH accretion, which are empirical relationships between BH mass, $M_\mathrm{BH}$, its nuclear X-ray luminosity at 2 - 10 keV, $L_\mathrm{X}$, and its nuclear radio luminosity at 5 GHz, $L_\mathrm{R}$ \citep{merl03}.  FPs cover six orders of magnitude in BH mass down to intermediate-mass black holes \citep[$< 10^5$ \Ms;][]{gul14}.  

\subsection{BH Radio Flux Density}

To find the BH radio flux density in a given band in the Earth frame we first use an FP to calculate $L_\mathrm{R}$ in the source frame, which depends on $M_\mathrm{BH}$ and $L_\mathrm{X}$.  We find $L_\mathrm{X}$ from $L_{\mathrm{bol}}$, the bolometric luminosity of the BH, with Equation 21 of \citet{marc04},
\begin{equation}
\mathrm{log}\left(\frac{L_\mathrm{bol}}{L_\mathrm{X}}\right) = 1.54 + 0.24 \mathcal{L} + 0.012 \mathcal{L}^2 - 0.0015 \mathcal{L}^3,
\end{equation}
where $L_\mathrm{bol}$ is in units of solar luminosity, $\mathcal{L} = \mathrm{log} \, L_\mathrm{bol} - 12$, and we take $L_\mathrm{bol} = 5 \times 10^{45}$ erg s$^{-1}$ from \citet{akos23}.  $L_\mathrm{R}$ can then be determined from $L_\mathrm{X}$ from the FP,
\begin{equation}
\mathrm{log} \, L_\mathrm{R} (\mathrm{erg \, s^{-1}})= A \, \mathrm{log} \, L_\mathrm{X} (\mathrm{erg \, s^{-1}}) + B \, \mathrm{log} \, M_\mathrm{BH} (\mathrm{M}_{\odot})+ C,
\end{equation}
where $A$, $B$ and $C$ are for FPs from \citet[][MER03]{merl03}, \citet[][KOR06]{kord06}, \citet[][GUL09]{gul09}, \citet[][PLT12]{plot12}, and \citet[][BON13]{bonchi13} and are listed in Table~1 of \citet{latif23a}.  We also include the FP of Equation 19 in \citet[][GUL19]{gul19}, which has a slightly different form:
\begin{equation}
R \, = \, -0.62 + 0.70 \, X + 0.74 \, \mu,
\end{equation}
where $R =$ log($L_\mathrm{R}/10^{38} \mathrm{erg/s}$), $X =$ log($L_\mathrm{X}/10^{40} \mathrm{erg/s}$) and $\mu =$ log($M_\mathrm{BH}/10^{8}$\Ms). 

Radio emission from a supermassive black hole (SMBH) that is cosmologically redshifted into a given observer band today does not originate from 5 GHz in the source frame at high redshifts, so we calculate it from $L_\mathrm{R} =$ $\nu L_{\nu}$ assuming that the spectral luminosity $L_{\nu} \propto \nu^{-\alpha}$.  We consider $\alpha =$ 0.7 and 0.3 to bracket a reasonble range of spectral profiles \citep{ccb02,glou21}.  The spectral flux at $\nu$ in the Earth frame is then calculated from the spectral luminosity at $\nu'$ in the rest frame from
\begin{equation}
F_\nu = \frac{L_{\nu'}(1 + z)}{4 \pi {d_\mathrm L}^2},
\end{equation}
where $\nu' = (1+z) \nu$ and $d_\mathrm L$ is the luminosity distance.  In calculating $d_\mathrm L$ we assume second-year \textit{Planck} cosmological parameters:  $\Omega_{\mathrm M} = 0.308$, $\Omega_\Lambda = 0.691$, $\Omega_{\mathrm b}h^2 = 0.0223$, $\sigma_8 =$ 0.816, $h = $ 0.677 and $n =$ 0.968 \citep{planck2}.
 
\subsection{Supernova Radio Emission}

Synchrotron emission from young SN remnants in the host galaxy could contribute to the total radio emission from UHZ1.  \citet{con92} estimate the non-thermal radio flux density from SNe in star-forming galaxies today to be
\begin{equation}
\left(\frac{L_{\mathrm{N}}}{\mathrm{W \, Hz^{-1}}}\right) \, \sim \, 4.4 \times 10^{34} \left(\frac{\nu}{\mathrm{GHz}}\right)^{-\beta} \left[\frac{\mathrm{SFR}(M > 5 \, \mathrm{M_{\odot}})}{\mathrm{M_{\odot}} \, \mathrm{yr^{-1}}} \right],
\label{eq:SNe}
\end{equation}
where $\beta \sim$ 0.8 is the nonthermal spectral index and we set $\mathrm{SFR} = 4 \, \Ms \, \mathrm{yr}^{-1}$ from \citet{akos23} for simplicity because the stellar IMF is unknown.  Equation~\ref{eq:SNe} may overstimate the SN flux density from UHZ1 because core-collapse SNe \citep{jet09b,latif22a} only produce nJy radio flux densities \citep{mw12} in the diffuse H II regions of high-redshift halos \citep{wan04,wet08a}.  Since we also assume that SF only produces stars $>$ 5 \Ms, this flux density should be taken to be an upper limit.

\subsection{H II Region Radio Emission}

Thermal bremsstrahlung in H II regions due to active SF also produces continuum radio emission whose spectral density can be calculated from the ionizing photon emission rate in the H II region, $Q_{\mathrm{Lyc}}$:
\begin{equation}
L_{\nu} \, \lesssim \, \left(\frac{Q_{\mathrm{Lyc}}}{6.3 \times 10^{52} \, \mathrm{s^{-1}}}\right) \left(\frac{T_{\mathrm{e}}}{10^4 \mathrm{K}}\right)^{0.45} \left(\frac{\nu}{\mathrm{GHz}}\right)^{-0.1}
\end{equation}
in units of 10$^{20}$ W Hz$^{-1}$ \citep{con92}, where $Q_{\mathrm{Lyc}} =$ SFR (\Ms \, yr$^{-1}$) $/$ $1.0 \times 10^{-53}$ \citep{ken98}.  We again take SFR $=$ 4 and $T_{\mathrm{e}} =$ 10$^4$ K. 

\section{Results and Discussion}

\begin{figure*} 
\begin{center}
\begin{tabular}{cc}
\includegraphics[width=0.42\linewidth]{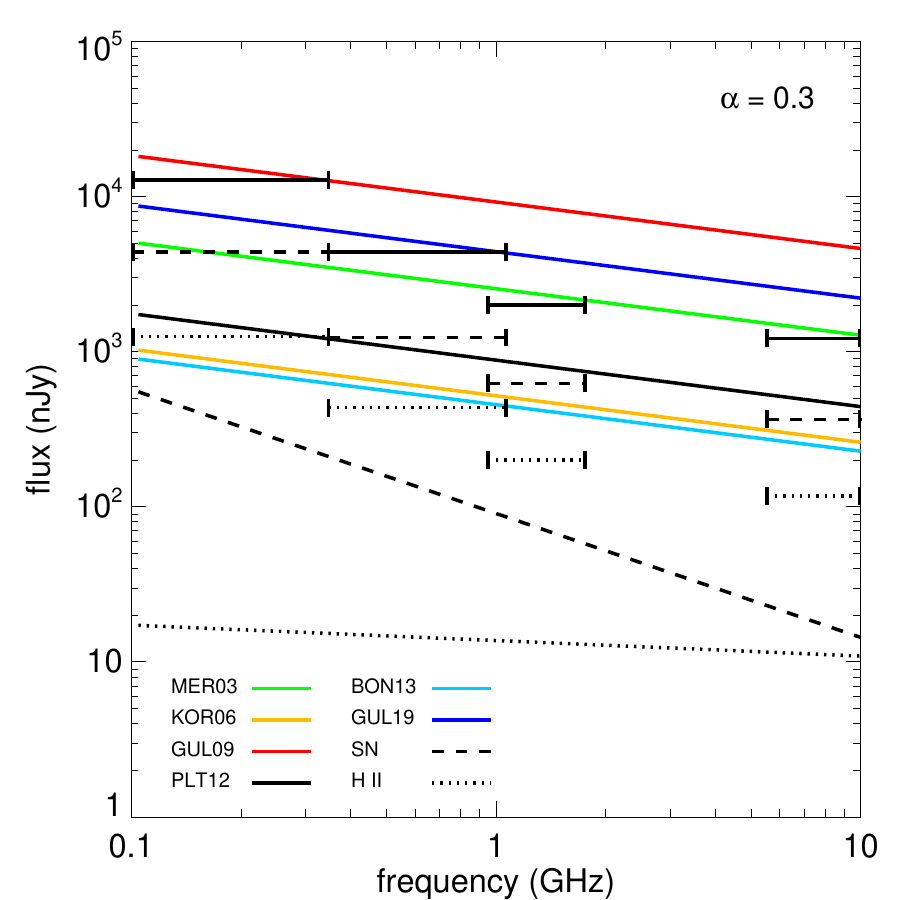}  &
\includegraphics[width=0.42\linewidth]{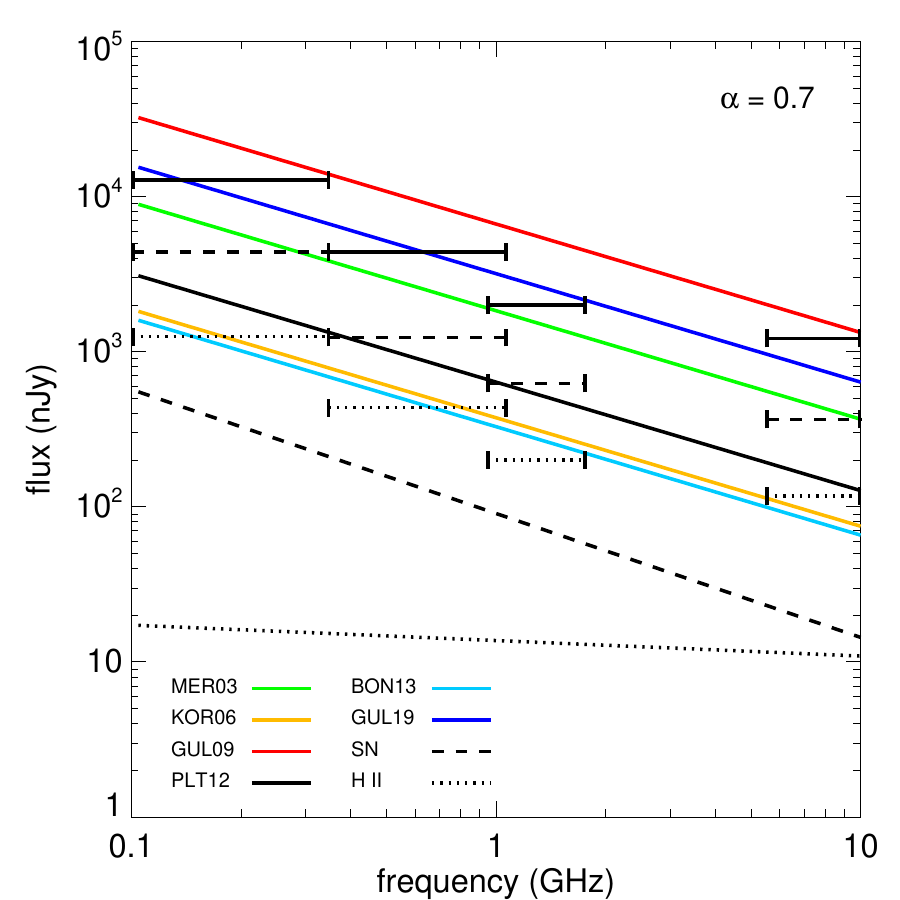}  \\
\includegraphics[width=0.42\linewidth]{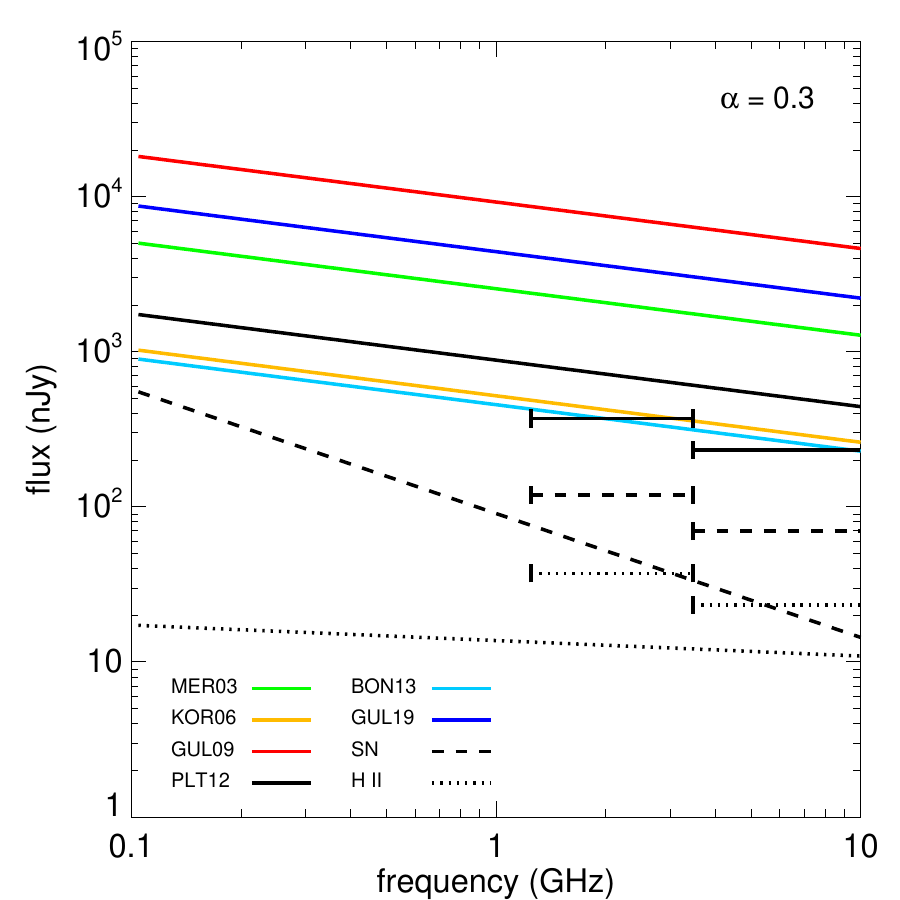}  &
\includegraphics[width=0.42\linewidth]{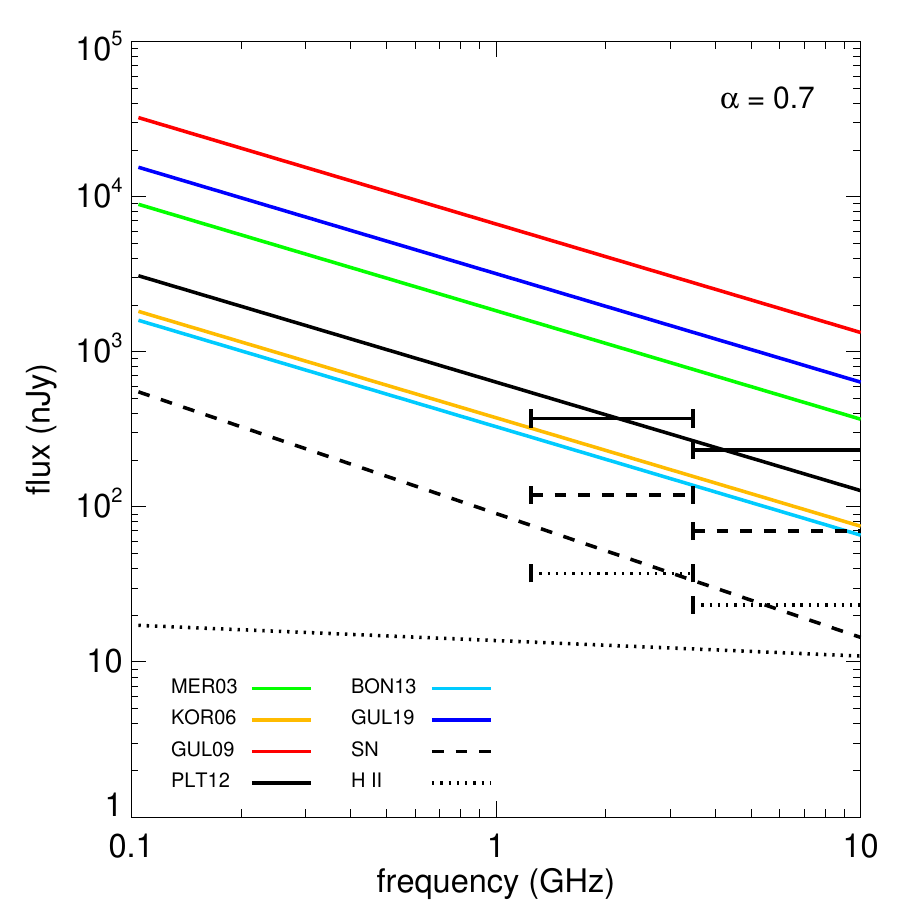}  
\end{tabular}
\end{center}
\caption{Radio flux densities for the BH in UHZ1 at 0.1 - 10 GHz.  The horizontal bars are detection limits as explained below.  Left panels:  $\alpha =$ 0.3.  Right panels: $\alpha =$ 0.7. Top row: horizontal lines indicate SKA detection limits at 0.1 - 0.35 GHz, 0.35 - 1.05 GHz, 0.95-1.76 GHz, and 4.6 - 10 GHz for 1 hr (solid), 10 hr (dashed) and 100 hr (dotted) integration times.  Bottom row: BH flux densities with ngVLA detection limits at 1.2 - 3.5 GHz and 3.5 - 10 GHz for 1 hr (solid), 10 hr (dashed) and 100 hr (dotted) integration times.  All flux densities have been boosted by the lensing factor $\mu =$ 3.81 from \citet{akos23}.}
\vspace{0.2in}
\label{fig:flux} 
\end{figure*}

We show radio flux densities for UHZ1 for $\alpha =$ 0.3 and 0.7 for all six FPs in Figure~\ref{fig:flux} along with the H II region and SN flux densities for its host galaxy.  They vary from 1.3 - 32 $\mu$Jy at 100 MHz to 0.065 - 1.3 $\mu$Jy at 10 GHz for $\alpha =$ 0.7 and from 0.9 - 20 $\mu$Jy at 100 MHz to 0.23 - 1.1 $\mu$Jy at 10 GHz for $\alpha =$ 0.3.  As expected, the highest densities are at the lowest frequencies, and they fall by a factor of a few to an order of magnitude by 10 GHz.  Most of the FP flux densities for $\alpha =$ 0.3 could be detected at 0.95 - 1.76 GHz and 4.6 - 10 GHz with 10 hr integration times by the SKA at \citep[see Table 3 of][]{ska} and virtually all of them could be found with 100 hr pointings, as shown by the detection bars in Figure~\ref{fig:flux}.  

The ngVLA could detect the smallest of the flux densities with just 1 hr integration times over its bandwidth, 1.2 - 3.5 GHz and 3.5 - 10 GHz \citep{pr18}.  For $\alpha =$ 0.7, the smallest of the flux densities could be detected with a 10 hr pointing by ngVLA and 100 hr integration times with SKA.  However, the top two FP flux densities could be found by just 1 hr integration times by SKA for both $\alpha$.  The H II region flux density is only $\sim$ 20 nJy but the SN flux density comes within a factor of 2 of the lowest BH flux densities at 0.1 GHz.  It falls steeply thereafter and is at most 1\% - 10\% of even the lowest BH flux densities above 1 GHz.  Consequently, any detection of radio emission from UHZ1 except perhaps at the lowest frequencies would unambiguously confirm the existence of a BH there.

UHZ1 does not appear in recent VLA or upgraded Giant Metrewave Radio Telescope (uGMRT) observations of Abel 2744 at the 7 $\mu$Jy/beam RMS noise level in the VLA $L$-band (1 - 2 GHz), at 5 $\mu$Jy in the VLA $S$-band (2 - 4 GHz), at 41 $\mu$Jy in uGMRT band 3 (300 - 500 MHz), or at 7 $\mu$Jy in uGMRT band 4 (550 - 950 MHz; Table 3 of \citealt{raj21} -- see also \citealt{pet17}).  Figure~\ref{fig:flux} shows that uGMRT therefore rules out the GUL09 and GUL19 FPs for $\alpha =$ 0.3 but only GUL09 for $\alpha =$ 0.7.  All the FPs are consistent with the VLA data because they fall below its noise limits.  

\section{Conclusion}

Our estimates show that radio emission from UHZ1 at $z =$ 10.3 would be easily detected by SKA and ngVLA in the coming decade, but with sufficient integration times it may be possible to detect with VLA now.  Radio measurements of UHZ1 might also detect features of the BH that may not appear in the NIR or X-rays, such as jets, which we do not consider here.  The angular diameter of UHZ1 is $\sim$ 0.25$^{\prime\prime}$ but SKA will reach angular resolutions of 0.08$^{\prime\prime}$ and 0.04$^{\prime\prime}$ at 6.5 GHz and 10 GHz, respectively, and could partially resolve radio structure originating from the host galaxy.  It is not clear if the BH in UHZ1 would have a jet because it is assumed to be accreting at about the Eddington limit and jets have mostly been observed at $L_{\mathrm{bol}} \lesssim 0.01 \, L_{\mathrm{Edd}}$ and $L_{\mathrm{bol}} \gtrsim L_{\mathrm{Edd}}$.  If a jet did form, the cosmic microwave background might quench its emission \citep{gh14,fg14}.  Nevertheless, compact jets have been observed from high-redshift quasars on scales of a kpc or less in one or two cases \citep{mom18,con21} and could be resolved by SKA or ngVLA measurements.

\citet{latif23a} found that SKA and ngVLA could detect BHs like UHZ1 at earlier stages of evolution at lower masses and higher redshifts, in principle up to $z \sim$ 14 - 16, but in targeted searches rather than blind surveys because the numbers of massive BHs at these redshifts are so small.  UHZ1 is an excellent such target, not only to complement prior NIR and X-ray measurements but to develop radio followups on photometric detections of SMBH candidates by {\em Euclid} and the {\em Roman Space Telescope} ({\em RST}) at $z \lesssim$ 15 in the coming decade.  The discovery of radio emission from UHZ1 would be the first test of potential synergies between NIR surveys and radio observations in detections of $z \sim$ 5 - 15 quasars in the coming decade.

\acknowledgments

The authors thank the referee, whose comments improved the quality of this paper, and Akos Bogdan, David Bacon and Matt Jarvis for helpful discussions.  MAL thanks the UAEU for funding via UPAR grants No. 31S390 and 12S111.  This work was also supported by the program Unidad de Excelencia Mar\'ia de Maeztu CEX2020-001058-M.  

\bibliographystyle{apj}
\bibliography{refs}

\end{document}